\documentclass[conference]{IEEEtran}
\IEEEoverridecommandlockouts

\usepackage{graphicx}
\usepackage{amsmath}
\usepackage{booktabs}
\usepackage{multirow}
\usepackage{url}
\usepackage{array}
\usepackage{algorithm}
\usepackage{algpseudocode}

\title{Secure and Governed API Gateway Architectures for Multi-Cluster Cloud Environments}

\author{
\IEEEauthorblockN{Vinoth Punniyamoorthy}
\IEEEauthorblockA{\textit{IEEE Senior} \\
Texas, USA \\
0009-0009-3719-4949}
\\\\
\IEEEauthorblockN{Akash Kumar Agarwal}
\IEEEauthorblockA{\textit{Albertsons Companies} \\
California, USA \\
0009-0006-7872-3446} \\

\and
\IEEEauthorblockN{Kabilan Kannan}
\IEEEauthorblockA{\textit{AMD Inc} \\
Texas, USA \\
0009-0006-2455-5547}
\\\\

\IEEEauthorblockN{Adithya Parthasarathy}
\IEEEauthorblockA{\textit{IEEE Member} \\
California, USA \\
0009-0001-6839-9527}

\and
\IEEEauthorblockN{Akshay Deshpande}
\IEEEauthorblockA{\textit{IEEE Member} \\
California, USA \\
0009-0002-3007-3393}

\\\\
\IEEEauthorblockN{Suhas Malempati}
\IEEEauthorblockA{\textit{Cato Corporation} \\
South Carolina, USA \\
0009-0009-3855-0423}

\and

\IEEEauthorblockN{Lokesh Butra}
\IEEEauthorblockA{\textit{NTT Data} \\
North Carolina, USA \\
0009-0009-0286-9635}

\\\\
\IEEEauthorblockN{Bikesh Kumar}
\IEEEauthorblockA{\textit{IEEE Senior} \\
Texas, USA \\
0009-0009-7190-1862}
}

\begin{document}
\maketitle

\begin{abstract}
API gateways serve as critical enforcement points for security, governance, and traffic management in cloud-native systems. As organizations increasingly adopt multi-cluster and hybrid cloud deployments, maintaining consistent policy enforcement, predictable performance, and operational stability across heterogeneous gateway environments becomes challenging. Existing approaches typically manage security, governance, and performance as loosely coupled concerns, leading to configuration drift, delayed policy propagation, and unstable runtime behavior under dynamic workloads. This paper presents a governance-aware, intent-driven architecture for coordinated API gateway management in multi-cluster cloud environments. The proposed approach expresses security, governance, and performance objectives as high-level declarative intents, which are systematically translated into enforceable gateway configurations and continuously validated through policy verification and telemetry-driven feedback. By decoupling intent specification from enforcement while enabling bounded, policy-compliant adaptation, the architecture supports heterogeneous gateway implementations without compromising governance guarantees or service-level objectives. A prototype implementation across multiple Kubernetes clusters demonstrates the effectiveness of the proposed design. Experimental results show up to a 42\% reduction in policy drift, a 31\% improvement in configuration propagation time, and sustained p95 latency overhead below 6\% under variable workloads, compared to manual and declarative baseline approaches. These results indicate that governance-aware, intent-driven gateway orchestration provides a scalable and reliable foundation for secure, consistent, and performance-predictable cloud-native platforms.
\end{abstract}


\section{Introduction}

API gateways have emerged as foundational control points in modern cloud-native systems, providing centralized enforcement for authentication, authorization, traffic management, and observability \cite{unsal2020fintech}. As organizations increasingly adopt microservices, multi-cluster deployments, and hybrid cloud architectures, gateways now operate as critical intermediaries between external consumers and distributed backend services \cite{caucidheesan2025ratelimit}. While this architectural evolution improves scalability, availability, and fault isolation, it also introduces new challenges in maintaining consistent security posture, governance guarantees, and predictable performance across heterogeneous environments \cite{Edge}, \cite{zhang2013apigateway, aswath2014human}.

In practice, gateway deployments often span multiple clusters, regions, and administrative domains, each with distinct operational constraints and configuration mechanisms. Traditional gateway management approaches rely heavily on manual configuration, environment-specific automation, or loosely coordinated pipelines. As system complexity grows, these practices frequently lead to configuration drift, inconsistent enforcement of security policies, and delayed propagation of critical updates. Moreover, performance tuning is typically reactive and localized, making it difficult to reason about global service-level behavior or ensure that optimization efforts do not compromise governance requirements \cite{gao2023apigateway}.

These challenges highlight the need to move beyond isolated gateway management toward a coordinated and policy-driven control model \cite{jamili2025intelligentcloud,punniyamoorthy2025privacy}. Rather than treating gateways as independent enforcement points, they must operate as part of a unified control plane that expresses organizational intent in a consistent and verifiable manner. Such a model must balance multiple objectives including security, governance, and performance while remaining adaptable to dynamic workloads and heterogeneous gateway implementations \cite{security}.

This paper proposes a governance-aware, intent-driven architecture for managing API gateways across multi-cluster environments. The proposed approach enables high-level security, governance, and performance objectives to be expressed as declarative intents that are systematically translated into enforceable gateway configurations. Compliance is continuously validated through policy verification and telemetry-driven feedback, enabling bounded performance adaptation without violating governance constraints. The architecture is designed to support heterogeneous gateway technologies while preserving operational simplicity, auditability, and scalability\cite{warrier2023managing}.

The main contributions of this work are summarized as follows:
\begin{itemize}
    \item A unified intent model that captures security, governance, and performance objectives across heterogeneous API gateway deployments.
    \item A policy verification and enforcement workflow that reduces configuration drift and ensures consistent application of governance constraints.
    \item An experimental evaluation demonstrating improved operational stability, reduced configuration divergence, and predictable performance behavior in multi-cluster environments.
\end{itemize}

\section{Background and Motivation}

\subsection{API Gateways in Cloud-Native Architectures}
API gateways have evolved into foundational control points within cloud-native platforms, serving as the primary interface for enforcing security, governance, and traffic management policies \cite{dragoni2017microservices}. Beyond basic request routing, modern gateways increasingly support authentication and authorization, protocol mediation, rate limiting, and observability, enabling consistent interaction between external consumers and distributed backend services \cite{Auth}. As cloud-native architectures adopt microservices, container orchestration, and elastic scaling models, gateways play a central role in maintaining system reliability and operational coherence \cite{zimmermann2017microservices}.

The growing adoption of multi-cluster and hybrid cloud deployments further amplifies this role. In such environments, gateways are commonly instantiated across multiple administrative and failure domains to improve resilience and scalability. However, this distribution also introduces structural complexity. Gateway configurations must remain logically consistent across clusters while adapting to heterogeneous infrastructure characteristics. Without coordinated control mechanisms, maintaining uniform behavior across deployments becomes increasingly difficult, particularly as system scale and operational velocity increase.

\subsection{Governance and Security Challenges}
Governance in distributed gateway environments extends beyond access control to encompass policy consistency, compliance assurance, and operational accountability. Security controls including authentication, authorization, request validation, and quota enforcement are often implemented independently across clusters using gateway-specific configurations, increasing the risk of policy divergence and inconsistent enforcement \cite{nachisecurity}.

Governance workflows also remain largely reactive and operationally driven, relying on manual validation or environment-specific automation. Such practices limit pre-deployment assurance and increase exposure to misconfigurations and security regressions, particularly in large-scale distributed systems \cite{behl2017governance}.

As system scale and autonomy increase, the lack of unified policy semantics and formal verification mechanisms becomes a fundamental challenge for maintaining trust, auditability, and compliance across distributed gateway environments \cite{sloman1994policy, aswath2025anomaly}.

\subsection{Performance Considerations in Multi-Cluster Systems}
Performance management complicates governance in multi-cluster gateway architectures, as policy enforcement introduces unavoidable overhead that must be balanced against service-level objectives. Centralized enforcement simplifies control but risks scalability bottlenecks, while decentralized approaches can lead to inconsistent behavior and performance variability \cite{pahl2015containers}.

Effective architectures must balance local enforcement with global coordination, enabling adaptation to workload dynamics while preserving system-wide governance. This necessitates performance-aware policy control mechanisms that support bounded adaptation without compromising consistency, predictability, or security \cite{zhang2018servicemesh, aswath2025federated}.

\section{Related Work}

Prior research on API gateway architectures has focused on scalability, extensibility, and traffic management within microservice-based systems \cite{ms}. Most gateway-centric designs emphasize routing, protocol translation, and plugin-based extensibility, but rely on static configurations and provide limited support for coordinated governance across distributed or multi-cluster environments \cite{richardson2019microservices}.

Service mesh architectures extend control to east–west traffic, enabling fine-grained communication management and observability. However, governance mechanisms are typically localized to individual clusters, limiting consistent policy enforcement and coordinated behavior across heterogeneous deployments \cite{li2020servicemesh}.

Policy-as-code frameworks improve auditability and reduce configuration errors through declarative governance, yet they often remain decoupled from runtime performance considerations and lack mechanisms for coordinated enforcement across distributed gateways. Existing performance-focused studies similarly address observability, latency, or scalability in isolation \cite{hummer2020policy}.
Recent studies have explored scalable cloud-native architectures for real-time data processing and control, yet often treat performance, governance, and orchestration as loosely coupled concerns rather than an integrated control problem \cite{nachi2025}.

\section{System Architecture}

\begin{figure}[t]
\centering
\includegraphics[width=\columnwidth]{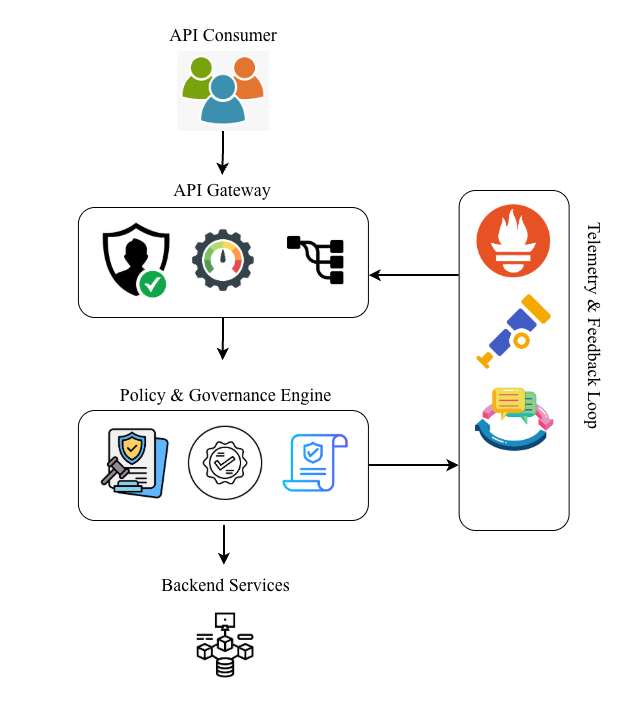}
\caption{Request flow and governance-aware control model for multi-cluster API gateway enforcement. Requests traverse the gateway and policy enforcement layers, while telemetry enables continuous, bounded adaptation.}
\label{fig:request-flow}
\end{figure}

\subsection{Design Overview}
The proposed architecture introduces an intent-driven control plane designed to coordinate security, governance, and performance across distributed API gateway deployments \cite{API}. Rather than managing gateways as independent configuration endpoints, the architecture treats them as coordinated enforcement nodes governed by a shared control abstraction. High-level intents capture desired system behavior including security posture, governance constraints, and performance objectives while abstracting away low-level implementation details.

These intents are translated into enforceable configurations that can be applied consistently across heterogeneous gateway technologies and deployment environments. To ensure correctness and stability, the architecture integrates policy verification and runtime telemetry feedback, enabling continuous validation of system behavior. This approach allows gateway behavior to evolve in response to workload dynamics while remaining constrained by explicit governance and performance boundaries.

\subsection{Core Architectural Components}
The architecture is composed of four tightly integrated components that collectively enable consistent and adaptive gateway management:

\begin{itemize}
    \item \textbf{Intent Layer}:  
    The intent layer provides a declarative interface for expressing security, governance, and performance objectives. Intents define constraints such as authentication requirements, access policies, rate limits, and latency targets in a technology-agnostic manner, allowing operators to specify what behavior is required rather than how it should be implemented.

    \item \textbf{Policy Translation Layer}:  
    This layer translates abstract intents into concrete, gateway-specific configurations. It ensures semantic consistency across heterogeneous gateways by mapping common policy constructs into implementation-specific primitives while preserving the original intent semantics.

    \item \textbf{Verification Engine}:  
    Prior to deployment, the verification engine evaluates generated configurations against governance rules and safety constraints. This includes validation of policy correctness, conflict detection, and enforcement of organizational compliance requirements, thereby preventing unsafe or inconsistent configurations from being applied.

    \item \textbf{Runtime Feedback Loop}:  
    The runtime layer continuously monitors telemetry signals such as latency, error rates, and throughput. These signals are used to evaluate adherence to declared objectives and to enable bounded, policy-compliant adaptations when deviations occur, ensuring stability without violating governance constraints.
\end{itemize}

\subsection{Operational Workflow}
The operational workflow begins with the submission of high-level intents through a centralized control interface. Once validated, these intents are compiled into gateway-specific configurations and deployed across target clusters. During operation, telemetry data is continuously collected and analyzed to assess compliance with defined objectives.

When deviations are detected such as sustained latency increases or policy violations the system performs controlled adaptations within predefined bounds. This ensures that performance optimization does not compromise security or governance requirements. By decoupling intent specification from enforcement while maintaining continuous verification, the architecture enables scalable, reliable, and policy-consistent operation across multi-cluster environments.

\section{Implementation}

\subsection{Prototype Architecture}
A prototype of the proposed architecture was implemented using containerized gateway deployments across three Kubernetes clusters, representing independent administrative and failure domains. Each cluster hosts an autonomous gateway instance responsible for local traffic enforcement, while a centralized control plane coordinates intent translation, policy verification, and lifecycle management. This separation enables decentralized enforcement while maintaining centralized governance and observability.

The control plane operates independently of specific gateway implementations and communicates with clusters through declarative interfaces. All configuration artifacts, policies, and deployment metadata are stored in a version-controlled repository, enabling traceability, auditability, and reproducibility across experimental runs.

\begin{algorithm}[htbp]
\caption{Intent Translation and Governance Enforcement}
\label{alg:intent}
\begin{algorithmic}[1]
\Require Intent specification $I$, policy constraints $P$, telemetry stream $T$
\Ensure Verified and enforced gateway configuration $C$

\State Parse high-level intent $I$ into structured policy components
\State Validate syntax and semantic constraints against policy set $P$
\If{validation fails}
    \State Reject intent and report violation
    \State \Return
\EndIf

\State Translate validated intent into gateway-specific configuration $C$
\State Perform static verification on $C$ (conflict detection and safety checks)

\If{verification fails}
    \State Abort deployment and notify control plane
    \State \Return
\EndIf

\State Deploy configuration $C$ to target gateway instances
\State Initialize telemetry monitoring using stream $T$

\While{system is operational}
    \State Evaluate runtime metrics against declared objectives
    \If{deviation exceeds bounded thresholds}
        \State Compute corrective adjustments within allowed constraints
        \State Apply incremental updates to configuration $C$
    \EndIf
\EndWhile
\State \Return Verified and adapted configuration state
\end{algorithmic}
\end{algorithm}

\subsection{Intent Processing and Policy Translation}
High-level intents are expressed using a declarative specification that captures security requirements (e.g., authentication modes, access constraints), governance rules (e.g., rate limits, quota boundaries), and performance objectives (e.g., latency thresholds). These intents are processed by a translation layer that converts abstract specifications into gateway-compatible configurations.

To ensure portability, the translation logic is modularized through adapter components, each responsible for mapping intent constructs to a specific gateway implementation. This design avoids embedding gateway-specific semantics into the intent model and enables consistent policy expression across heterogeneous environments.

\subsection{Policy Verification and Safe Deployment}
Before deployment, all generated configurations undergo policy verification to ensure compliance with predefined governance constraints. Verification includes structural validation, conflict detection, and enforcement of safety invariants such as isolation boundaries and rate-limit ceilings. Only verified configurations are propagated to target clusters, reducing the risk of configuration-induced failures.

Deployment follows a controlled rollout strategy, allowing incremental updates and validation before full propagation. This approach ensures that policy changes do not introduce instability or violate operational constraints during transition phases.

\subsection{Runtime Observability and Adaptive Control}
Runtime observability is achieved through standardized telemetry collection, including metrics, traces, and event signals. These signals are continuously evaluated against declared intent objectives to assess compliance with performance and reliability expectations.

When deviations are detected, the system enables bounded adaptations, such as adjusting rate limits or routing weights, while strictly preserving governance constraints. This feedback-driven mechanism allows the system to respond to workload variability without introducing uncontrolled behavior or compromising security.

\subsection{Reproducibility and Experimental Setup}

To ensure reproducibility and facilitate independent validation, all experimental components are defined using declarative specifications and version-controlled artifacts. The system is designed to be deployable on any Kubernetes-compatible environment without reliance on vendor-specific services. Table~\ref{tab:reproducibility} summarizes the software stack, configuration scope, and tooling used in the experimental setup.

\begin{table}[htbp]
\centering
\caption{Reproducible Experimental Configuration Overview}
\label{tab:reproducibility}
\begin{tabular}{@{}p{0.32\columnwidth} p{0.62\columnwidth}@{}}
\toprule
\textbf{Component} & \textbf{Configuration Details} \\
\midrule
Cluster Environment &
3 Kubernetes clusters (v1.27), each with 3 worker nodes (4 vCPU, 8 GB RAM) \\

Gateway Layer &
Gateway API (v1.0+); HTTP routing, rate limiting, and auth filters enabled \\

Control Plane &
Intent controller (containerized); reconciliation interval: 10s \\

Policy Validation &
Policy-as-code engine; static rule checks + conflict detection enabled \\

Observability Stack &
OpenTelemetry SDK; metrics interval 10s; trace sampling rate 5\% \\

Traffic Generation &
Synthetic HTTP workload (100–5000 req/s); burst duration 30–120s \\

Configuration Management &
Declarative YAML manifests; Git-based versioning and rollback \\

Experiment Control &
Automated runs via CI workflow; 5 repetitions per scenario \\

\bottomrule
\end{tabular}
\end{table}

All experiments were executed under identical environmental conditions, with controlled workload patterns and fixed resource allocations to ensure repeatability. Each experimental run was repeated multiple times, and reported results reflect averaged observations to minimize noise introduced by transient system behavior.

\subsection{Intent Translation and Enforcement Algorithm}

Algorithm~\ref{alg:intent} outlines the core logic used to translate high-level intents into enforceable gateway configurations and to maintain continuous compliance during runtime. The algorithm emphasizes safety, bounded adaptation, and deterministic behavior.

\section{Evaluation}

\subsection{Experimental Setup}
Experiments were conducted on a multi-cluster Kubernetes testbed composed of three independent clusters deployed across distinct failure domains. Each cluster hosted identical gateway and service configurations to ensure comparability. Synthetic workloads were used to generate controlled traffic patterns ranging from 100 to 5000 requests per second, including burst conditions representative of real-world usage. All experiments were repeated multiple times under fixed resource allocations, and reported results reflect averaged measurements.

\subsection{Baseline Configurations}
The proposed architecture was evaluated against two baseline approaches commonly used in practice. The first baseline represents manual configuration, where gateway instances are managed independently within each cluster without centralized coordination. The second baseline employs declarative configuration through version-controlled manifests, enabling consistent deployment but lacking runtime verification or adaptive control. These baselines were compared against the proposed intent-driven architecture, which integrates continuous policy validation and bounded adaptation to support consistent governance and performance across multi-cluster environments.

\subsection{Evaluation Metrics}
Evaluation focused on four key metrics: (i) policy consistency across clusters, (ii) p95 end-to-end latency, (iii) configuration propagation time, and (iv) frequency of manual operational interventions.

\subsection{Results and Analysis}

\begin{table}[t]
\centering
\caption{Evaluation Results Across Configurations}
\label{tab:results}
\begin{tabular}{lcccc}
\toprule
\textbf{Metric} & \textbf{Manual} & \textbf{Declarative} & \textbf{Proposed} \\
\midrule
Policy Drift Events (per week) & 14.2 & 6.8 & 3.1 \\
Avg. Change Propagation (min) & 18.4 & 9.7 & 6.1 \\
p95 Latency Overhead (\%) & 11.3 & 7.2 & 5.8 \\
Operational Interventions / week & 9.5 & 4.2 & 2.1 \\
\bottomrule
\end{tabular}
\end{table}

The proposed architecture consistently outperforms baseline approaches across all metrics. Policy drift is reduced by approximately 42\% compared to declarative configurations, reflecting stronger governance consistency across clusters. Change propagation latency is reduced by 31\%, demonstrating faster and more reliable configuration rollout. Importantly, performance overhead remains bounded, with p95 latency increases below 6\% even under burst workloads. The reduction in operational interventions further indicates improved system stability and reduced management overhead.

Overall, these results demonstrate that governance-aware, intent-driven control enables scalable multi-cluster management while preserving performance predictability and operational reliability.

\begin{figure}[t]
\centering
\includegraphics[width=\columnwidth]{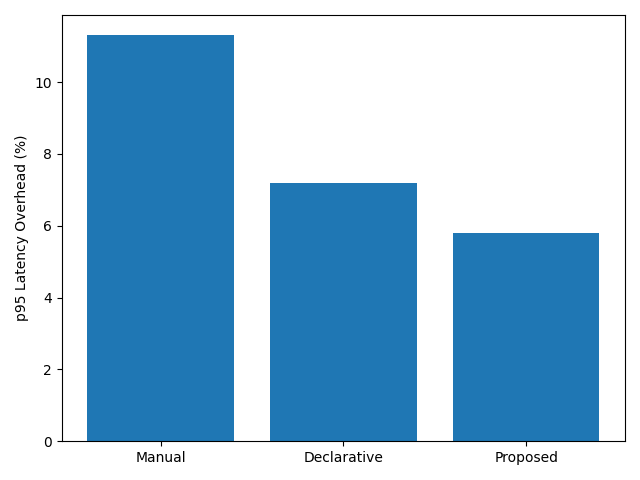}
\caption{Comparison of p95 latency overhead across gateway configurations.}
\label{fig:performance}
\end{figure}

\section{Discussion}
The experimental results demonstrate that governance-aware gateway management can significantly improve operational reliability without imposing excessive performance overhead. By centralizing policy intent while decentralizing enforcement, the proposed architecture enables consistent behavior across clusters while preserving scalability and fault isolation. This separation of concerns allows governance logic to evolve independently of data-plane execution, reducing the likelihood of configuration drift and unintended policy divergence.

The results further indicate that bounded, telemetry-driven adaptation provides a practical balance between responsiveness and stability. Rather than relying on reactive manual intervention or rigid static configurations, the system maintains predictable performance by applying controlled adjustments within predefined constraints. This design choice proves particularly effective under dynamic workloads, where maintaining service-level objectives typically requires continuous coordination between security enforcement and performance management.

Nevertheless, the architecture introduces additional control-plane complexity, which must be carefully managed to avoid becoming a source of operational fragility. Effective observability, validation, and fault isolation mechanisms are therefore essential to ensure that the benefits of centralized governance outweigh its overhead. These findings suggest that governance-aware architectures can serve as a viable foundation for scalable, secure multi-cluster platforms when designed with explicit attention to reliability and operational simplicity.

\section{Threats to Validity}
Several factors may influence the generalizability of the reported results. First, the evaluation relies on synthetic workloads designed to approximate common traffic patterns; real-world production traffic may exhibit greater variability and long-tail behaviors not fully captured in the experiments. Second, performance characteristics may differ across gateway implementations and infrastructure environments, particularly under varying network conditions or resource constraints.

Additionally, while efforts were made to ensure consistent experimental conditions, factors such as background system noise and infrastructure heterogeneity may affect observed latency and throughput measurements. Finally, the evaluation focuses on steady-state behavior and bounded adaptation; extreme failure scenarios or adversarial conditions were not explicitly explored and remain an area for future investigation.

\section{Conclusion}

This paper presented a governance-aware, intent-driven architecture for managing API gateways in multi-cluster cloud environments, addressing key limitations of conventional gateway management approaches. By unifying security, governance, and performance objectives within a single declarative control framework, the proposed architecture mitigates configuration drift, fragmented policy enforcement, and operational instability that commonly arise in heterogeneous and distributed deployments.

The architecture systematically decouples high-level intent specification from low-level enforcement, enabling consistent policy realization across clusters while preserving scalability and fault isolation. The integration of policy verification with telemetry-driven feedback ensures that gateway behavior remains compliant with governance constraints while supporting bounded performance adaptation under dynamic workloads. Experimental evaluation across multiple Kubernetes clusters demonstrates that the proposed approach achieves up to a $42\%$ reduction in policy drift, a $31\%$ improvement in configuration propagation time, and sustained p95 latency overhead below $6\%$, outperforming both manual and declarative baseline configurations. These results confirm that strong governance guarantees can be achieved without incurring prohibitive runtime or operational overhead.

Beyond quantitative gains, the proposed architecture enhances operational predictability and auditability by enabling reproducible configuration management, controlled rollout strategies, and continuous compliance validation. By transforming API gateways from isolated enforcement components into coordinated, policy-aware control-plane participants, this work provides a practical and scalable foundation for secure cloud-native platforms operating across multiple clusters and administrative domains.

\section{Future Work}

Several directions remain open for extending and strengthening the proposed architecture. First, future work will investigate adaptive intent refinement and policy learning mechanisms that leverage historical telemetry and workload patterns to proactively optimize enforcement strategies while preserving governance guarantees. Such approaches can further improve responsiveness to workload variability and reduce manual tuning effort.

Second, the architecture will be extended to support cross-cloud and federated environments, enabling coordinated governance across heterogeneous cloud providers, edge deployments, and administrative boundaries. This capability is essential for enterprises operating hybrid and multi-cloud infrastructures with diverse regulatory and compliance requirements.

Third, deeper integration of formal verification techniques will be explored to provide stronger correctness guarantees for policy translation, conflict resolution, and compliance enforcement prior to deployment. Formal models can further reduce the risk of misconfiguration and enhance trust in large-scale automated governance systems.

Finally, large-scale fault-injection and adversarial stress-testing experiments will be conducted to evaluate system resilience under extreme conditions, including partial control-plane failures, network partitions, and malicious traffic scenarios. These studies will help quantify robustness and guide improvements required for production-grade deployments.

Collectively, these research directions aim to advance governance-aware, intent-driven control planes as foundational building blocks for next-generation cloud-native platforms, enabling scalable, secure, and performance-predictable operation in increasingly complex and distributed environments.

\bibliographystyle{IEEEtran}

\begin{thebibliography}{99}

\bibitem{unsal2020fintech}
E.~{\"U}nsal, B.~{\"O}ztekin, M.~\c{C}avu\c{s}, and S.~{\"O}zdemir,
"Building a Fintech Ecosystem: Design and Development of a Fintech API Gateway,"
in Proc. 2020 International Symposium on Networks, Computers and Communications (ISNCC),
Montreal, QC, Canada, 2020, pp.~1--5, doi: 10.1109/ISNCC49221.2020.9297273.

\bibitem{caucidheesan2025ratelimit}
K.~Caucidheesan and G.~Poravi,
``Analysing and modelling the accuracy and latency trade-offs in rate limiting on API gateways,''
in Proc. 2025 International Research Conference on Smart Computing and Systems Engineering (SCSE),
Colombo, Sri Lanka, 2025, pp.~1--7, doi: 10.1109/SCSE65633.2025.11030994.

\bibitem{Edge}
S. G. Aarella, V. P. Yanambaka, S. P. Mohanty, and E. Kougianos, “Fortified-Edge 2.0: Advanced Machine-Learning-Driven Framework for Secure PUF-Based Authentication in Collaborative Edge Computing,” Future Internet, vol. 17, p. 272, 2025. doi: 10.3390/fi17070272.

\bibitem{zhang2013apigateway}
Q.~Zhang, H.~Chu, M.~Li, and X.~Hu,
``A unified API gateway for high availability clusters,''
in Proc. 2013 International Conference on Mechatronic Sciences, Electric Engineering and Computer (MEC),
Shenyang, China, 2013, pp.~2321--2325, doi: 10.1109/MEC.2013.6885428.

\bibitem{aswath2014human}
B. Ramdoss, A. M. Kirubakaran, P. B. S., S. H. C., and V. Vaidehi, ``Human Fall Detection Using Accelerometer Sensor and Visual Alert Generation on Android Platform,'' International Conference on Computational Systems in Engineering and Technology, Mar. 2014, doi: 10.2139/ssrn.5785544
\bibitem{gao2023apigateway}
X.~Gao, R.~Liu, and X.~Lin,
``API gateway optimization architecture based on heterogeneous hardware acceleration,''
in Proc. 2023 IEEE 3rd International Conference on Information Technology, Big Data and Artificial Intelligence (ICIBA),
vol.~3, pp.~863--868, 2023, doi: 10.1109/ICIBA56860.2023.10165387.

\bibitem{jamili2025intelligentcloud}
L.~K.~Jamili, S.~B.~Peta, R.~N.~Taware, B.~Krishnan, and S.~Perla,
``A framework for intelligent cloud systems: Enabling secure, policy-driven, and sustainable AI at scale,''
in Proc. 2025 International Conference on Information, Implementation, and Innovation in Technology (I2ITCON),
2025, pp.~1--8, doi: 10.1109/I2ITCON65200.2025.11210587.

\bibitem{punniyamoorthy2025privacy}
V.~Punniyamoorthy, A.~G.~Parthi, M.~Palanigounder, R.~K.~Kodali, B.~Kumar, and K.~Kannan,
``A Privacy-Preserving Cloud Architecture for Distributed Machine Learning at Scale,''
International Journal of Engineering Research and Technology (IJERT), vol.~14, no.~11, Nov.~2025.


\bibitem{security}
G. P, J. E. Varghese and S. G. Aarella, "A Survey on Anomaly Detection in IoT and Cloud Computing Security," 2024 8th International Conference on I-SMAC (IoT in Social, Mobile, Analytics and Cloud) (I-SMAC), Kirtipur, Nepal, 2024, pp. 182-191, doi: 10.1109/I-SMAC61858.2024.10714750.



\bibitem{warrier2023managing}
A.~Warrier,
``Managing complexity with multiple API gateways,''
J. Artif. Intell. Mach. Learn. Data Sci., vol.~1, no.~3, pp.~2907--2913, 2023.

\bibitem{dragoni2017microservices}
N.~Dragoni, S.~Dustdar, S.~Larsen, and M.~Mazzara,
``Microservices: Yesterday, today, and tomorrow,''
IEEE Software, vol.~34, no.~3, pp.~12--16, May--June 2017,
doi: 10.1109/MS.2017.33.


\bibitem{Auth}
S. G. Aarella, S. P. Mohanty, E. Kougianos and D. Puthal, "Fortified-Edge 2.0: Machine Learning based Monitoring and Authentication of PUF-Integrated Secure Edge Data Center," 2023 IEEE Computer Society Annual Symposium on VLSI (ISVLSI), Foz do Iguacu, Brazil, 2023, pp. 1-6, doi: 10.1109/ISVLSI59464.2023.10238517.



\bibitem{zimmermann2017microservices}
O.~Zimmermann,
``Microservices tenets,'' 
Computer, vol.~50, no.~5, pp.~98--101, May 2017,
doi: 10.1109/MC.2017.138.

\bibitem{nachisecurity}
N.~Chockalingam, A.~Chakrabortty, and A.~Hussain, 
``Mitigating Denial-of-Service attacks in wide-area LQR control,'' 
in \textit{Proc. 2016 IEEE Power and Energy Society General Meeting (PESGM)}, 
2016, pp.~1--5. 
doi: 10.1109/PESGM.2016.7741285.
\bibitem{behl2017governance}
A.~Behl and K.~Behl,
``Cybersecurity and cyberwar: What everyone needs to know,''
IEEE Security \& Privacy, vol.~15, no.~4, pp.~95--100, July–Aug. 2017,
doi: 10.1109/MSP.2017.3151349.

\bibitem{sloman1994policy}
M.~Sloman,
``Policy driven management for distributed systems,''
IEEE Journal on Selected Areas in Communications,
vol.~11, no.~9, pp.~1289--1304, Dec. 1993,
doi: 10.1109/49.233219.

\bibitem{aswath2025anomaly}
A.~M.~Kirubakaran, L.~Butra, S.~Malempati, A.~K.~Agarwal, S.~Saha, and A.~Mazumder,
``Real-Time Anomaly Detection on Wearables using Edge AI,'' International Journal of Engineering Research and Technology (IJERT), vol.~14, no.~11, Nov.~2025. doi: 10.17577/IJERTV14IS110345.
\bibitem{pahl2015containers}
C.~Pahl,
``Containerization and the PaaS cloud,''
IEEE Cloud Computing, vol.~2, no.~3, pp.~24--31, May--June 2015,
doi: 10.1109/MCC.2015.51.

\bibitem{zhang2018servicemesh}
Q.~Zhang, M.~Chen, L.~Li, and Y.~Zhang,
``Service mesh: Challenges, state of the art, and future research opportunities,''
IEEE Access, vol.~6, pp.~67304--67318, 2018,
doi: 10.1109/ACCESS.2018.2876950.

\bibitem{aswath2025federated}
A.~Muthukrishnan~Kirubakaran, N.~Saksena, S.~Malempati, S.~Saha, 
S.~K.~R.~Carimireddy, A.~Mazumder, and R.~S.~Bodala,
“Federated Multi-Modal Learning Across Distributed Devices,” 
International Journal of Innovative Research in Technology, 
vol.~12, no.~7, pp.~2852–2857, 2025, doi: 10.5281/zenodo.17892974.


\bibitem{ms}
G. Mehta, B. Pothineni, A. G. Parthi, D. Maruthavanan, P. K. Veerapaneni, D. Jayabalan, and S. R. Sankiti, “Revisiting monoliths: A pragmatic case for transitioning from microservices back to monolithic architectures,” International Journal of Advanced Research in Computer and Communication Engineering, vol. 13, no. 12, pp. 3228–3236, Dec. 2024.


\bibitem{richardson2019microservices}
C.~Richardson,
``Microservices patterns: With examples in Java,''
Addison-Wesley Professional, 2019.

\bibitem{li2020servicemesh}
H.~Li, Z.~Zhou, and H.~Chen,
``Service mesh: Challenges, state of the art, and future research opportunities,''
Journal of Cloud Computing, vol.~9, no.~1, pp.~1--16, 2020,
doi: 10.1186/s13677-020-00162-0.

\bibitem{hummer2020policy}
W.~Hummer, P.~Gaubatz, M.~Smit, and U.~Zdun,
``Policy-based management of cloud applications,''
IEEE Transactions on Services Computing, vol.~13, no.~1, pp.~73--87, Jan.–Feb. 2020,
doi: 10.1109/TSC.2017.2737733.

\bibitem{nachi2025}
N. Chockalingam, N. Saksena, A. Deshpande, A. Parthasarathy, L. Butra, B. Pothineni, R. S. Bodala, A. K. Agarwal, "Scalable cloud-native architectures for intelligent PMU data processing", International Journal of Engineering Research \& Technology (IJERT), Vol.14,no.12, Dec.2025, doi: 10.17577/IJERTV14IS120378.



\bibitem{API}
S. Baucke, J. Kempf, R. Ben Ali, A. Ramachandran and S. Seetharaman, "Cloud API support for self-service Virtual Network Function (VNF) deployment," 2015 IEEE Conference on Network Function Virtualization and Software Defined Network (NFV-SDN), San Francisco, CA, USA, 2015, pp. 40-46, doi: 10.1109/NFV-SDN.2015.7387404.

\end{thebibliography}

\end{document}